# Zero-point term and quantum effects in the Johnson noise of resistors: A critical appraisal

**Laszlo B. Kish [1], Gunnar A. Niklasson [2], Claes G. Granqvist [2]**

[1] Department of Electrical and Computer Engineering, Texas A&M University, TAMUS 3128, College Station, TX 77843-3128, USA
[2] Department of Engineering Sciences, The Ångström Laboratory, Uppsala University, P. O. Box 534, SE-75121 Uppsala, Sweden
E-mail: Laszlokish@tamu.edu, Gunnar.Niklasson@Angstrom.uu.se, Claes-Goran.Granqvist@angstrom.uu.se

**Abstract**. There is a longstanding debate about the zero-point term in the Johnson noise voltage of a resistor. This term originates from a quantum-theoretical treatment of the Fluctuation-Dissipation Theorem (FDT). Is the zero-point term really there, or is it only an experimental artefact, due to the uncertainty principle, for phase-sensitive amplifiers? Could it be removed by renormalization of theories? We discuss some historical measurement schemes that do not lead to the effect predicted by the FDT, and we analyse new features that emerge when the consequences of the zero-point term are measured via the mean energy and force in a capacitor shunting the resistor. If these measurements verify the existence of a zero-point term in the noise, then two types of perpetual motion machines can be constructed. Further investigation with the same approach shows that, in the quantum limit, the Johnson–Nyquist formula is also invalid under general conditions even though it is valid for a resistor-antenna system. Therefore we conclude that and a satisfactory quantum theory of the Johnson noise, the Fluctuation-Dissipation Theorem, must, as a minimum, include also the measurement system used to evaluate the observed quantities. Issues concerning the zero-point term may also have implications for phenomena in advanced nanotechnology.

## 1. Introduction

Thermal noise (Johnson noise) in resistors was discovered by Johnson [1] and explained by Nyquist [2] in 1927, one year after the foundations of quantum physics were completed. The Johnson–Nyquist formula states that

$$S_u(f) = 4R(f)hfN(f,T) \,, \tag{1}$$

where $S_u(f)$ is the one-sided power density spectrum of the voltage noise on the open-ended complex impedance $Z(f)$ with real part $\text{Re}[Z(f)] = R(f)$, and $h$ is Planck's constant. The Planck number $N(f,T)$ is the mean number of $hf$ energy quanta in a linear harmonic oscillator with resonance frequency $f$ at temperature $T$ and is given by

$$N(f,T) = \left[\exp(hf/kT) - 1\right]^{-1} . \tag{2}$$

Hence we have the well-known $N(f,T) = kT/(hf)$ case for the classical physical range with $kT \gg hf$. Eq. 2 results in an exponential cut-off for the Johnson noise in the quantum range with $f > f_P = kT/h$, in accordance with Planck's thermal radiation formula. In the deeply classical (low-frequency) limit, with $f \ll f_P = kT/h$, Eqs. 1 and 2 yield the familiar form used at low frequencies, *i.e.*,

$$S_{u,l}(f) = 4kTR(f) , \tag{3}$$

where the Planck cut-off frequency $f_P$ is about 6000 GHz at room temperature. This is well beyond the reach of today's electronics.

A quantum-theoretical treatment of the one-sided power density spectrum of the Johnson noise was given 24 years after Johnson's and Nyquist's work by Callen and Welton [3] (often referred to as the Fluctuation-Dissipation Theorem, FDT). The quantum version [3] of the Johnson–Nyquist formula has a number 0.5 added to the Planck number, corresponding to the zero-point (ZP) energy of linear harmonic oscillators, so that

$$S_{u,q}(f) = 4R(f)hf\left[N(f,T) + 0.5\right] . \tag{4}$$

Thus the quantum correction of Eq. 1 is a temperature-independent additive term in Callen–Welton's one-sided power density spectrum (Eq. 2) according to

$$S_{u,ZP}(f) = 2hfR(f) , \tag{5}$$

which depends linearly on frequency and exists for any $f > 0$, even in the deeply classical frequency regime and at zero temperature. The zero-point term described by Eq. 5 has gained widespread theoretical support over the years [4-8].

We note that absolute zero temperature cannot be reached in a physical system which means that, when discussing the zero-temperature limit, we always assume a non-zero temperature that is close-enough to zero so that $N(f,T) \ll 0.5$ holds at the measurement frequency.

We emphasize that Callen–Welton's derivation works solely with the *one-sided* spectrum, while subsequent quantum-theoretical approaches often utilize asymmetrical power density spectra of fluctuations [7,8] and are in full agreement with the Callen–Welton result.

## 2. The debate

The zero-point energy in the Johnson noise has been the subject of much discussion for many years. Without the goal of completeness, we briefly survey the most important arguments below.

2.1. The ground state

MacDonald [9] and Harris [10] claimed that extracting energy/power from the zero-point energy is impossible, and thus Eq. 5 should not exist.

2.2. Planck's black-body radiation

Grau and Kleen [11] and Kleen [12] (as in the original treatment by Nyquist [2]), argued that the Johnson noise of a resistor connected to an antenna, see Fig. 1, must satisfy Planck's thermal radiation formula. Thus the noise must be zero at zero temperature, which implies that Eqs. 4 and 5 are invalid. This argument is obvious even for a naked-eye observer: At a temperature of 6000 K (corresponding to a wavelength of ~600 nm, $i.e.$, orange-colored light), the Planck number is $N = 0.0164$. Thus the zero-point term in Eq. 4 is thirty times greater than the classical term, yet it is invisible to the eye and to a photocell and any other photon-counting detector.

Defenders of Eq. 4 may (incorrectly) say that the same zero-point term exists also in the thermal radiation field, which means that the net energy flow between the resistor and the radiation field is null for the zero-point term, just as it is for the classical term, by satisfying the Second Law of Thermodynamics. However, this argument fails when confronted with an objection based on fluctuations, even if we neglect the obvious problem that photon absorption in a photocell is irreversible. The zero-point terms in Eqs. 4 and 5 represent noises, and that means statistical fluctuations [12,13] of their finite-time mean-square values. The implication is that, for independent zero-point noises in the resistor and the radiation field, a "zero-point energy flow" with fluctuating direction and value of the short-time average should be observable between the antenna's input and its radiation field. But this is not the case, and it is a hard experimental fact that neither the zero-point term nor its fluctuations are observable in thermal radiation.

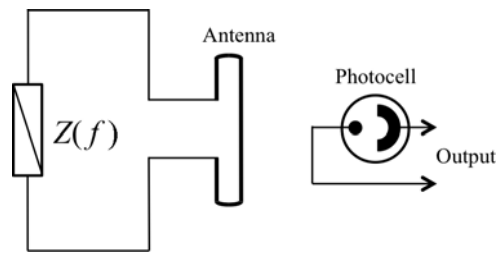

**Figure 1.** Measurement scheme, based on an antenna and a photon counter, which does not show the presence of the zero-point term (Eq. 5) or its fluctuation in the Johnson noise at its output [11,12].

2.3. Divergent noise energy

Kish [14] has pointed out that the existence of the zero-point term, which has an "$f$"-noise spectrum, implies 1/$f$ noise and a related logarithmic divergence of the energy in a shunt capacitor in the high-frequency limit. While this does not disprove the existence of Eq. 5, it may indicate that there is a renormalization problem—$i.e.$, a mathematical artifact—producing an unphysical term that is not actually present in measurements (in analogy with renormalization problems of ground states in quantum electrodynamics). Later Abbott $et al.$ [13] arrived at a different but unclear conclusion that "zero-point energy is infinite thus it should be renormalized but not the 'zero-point fluctuations'".

2.4. A crucial experimental proof?

Notwithstanding the criticisms mentioned above, an experiment by Koch, van Harlingen and Clarke [15] confirmed the validity of Eqs. 4 and 5 by measurements on resistively shunted Josephson junctions by use of a heterodyne measurement method (as necessitated by the high frequency); see Fig. 2. The scheme is taken to be equivalent to the standard linear amplifier/filter method that determines the one-sided power density spectrum of the noise but allows accessing very high frequencies.

2.5. The uncertainty-principle argument

Haus [16] and Kleen [17] used Heffner's theory [18], based on the time-energy uncertainty principle of frequency/phase selective linear amplifiers, and stated that the zero-point term in Eqs. 4 and 5 is a direct consequence of the *energy-time uncertainty principle* for phase-sensitive amplitude measurements (Fig. 2). The same argumentation implies that the antenna arrangement [11,12] (Fig. 1) will not show any uncertainty (and zero-point term) in the photon number. However, the uncertainty principle argument *cannot* disprove Eqs. 4 and 5. The asserted zero-point term in the noise voltage may still exist and also satisfy the uncertainty principle instead of being solely an experimental artifact.

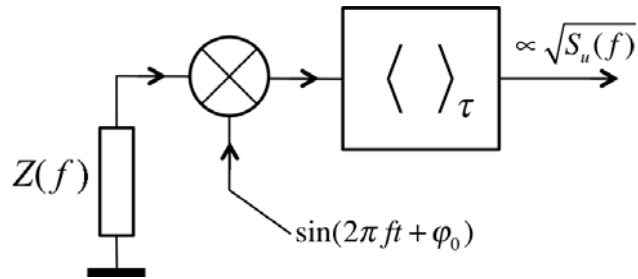

**Figure 2.** Heterodyne measurement scheme [15] based on a Josephson junction that mixes down the noise in the frequency range of interest. The non-linear mixer is represented by an analog multiplier symbol and is driven by the noise at one of its inputs and by the sinusoidal voltage oscillation at the Josephson frequency, $f = 2qU_{dc}/h$, at its other input, where $q$ is a charge quantum and $U_{dc}$ is the dc voltage on the Josephson junction. The dc component of the down-converted noise is proportional to $S_u^{0.5}(f)$ and is extracted by a time-average unit with a time constant $\tau$. Other filters and devices are not shown.

2.6. Criticism of the Callen–Welton theory

Recently, Reggiani *et al.* [19] generalized the derivation of Eq. 4 by including a discrete eigenvalue spectrum of the physical system of interest. They proposed (see Eq. 8 in their paper [19]) how the Callen–Welton relation, and the associated zero-point contribution to the noise spectrum, should be modified in this new context. Their result is interesting but not easy to interpret as regards the existence or non-existence of the zero-point term.

**3. A new approach to assess zero-point Johnson noise: Energy and force in a capacitor**

For the sake of simplicity, we assume that the resistors and capacitors discussed in the rest of the paper are macroscopic with sufficiently large density of defects that yield strong-enough defect scattering so that the phase breaking length [20] of charge transport is always much less than the smallest characteristic size of the resistors and capacitor. Thus the resistance does not converge to zero but saturates at a nonzero, low-temperature residual value (effect used, for example, in low-temperature noise-thermometry). This assumption does not reduce the significance of our results and claims because the Second Law of Thermodynamics must be valid at arbitrary conditions in thermal equilibrium.

For our present considerations of the zero-point term in the Johnson noise, the main conclusion of the debates outlined above is that the actual measurement scheme has a crucial role in the outcome of the observation. Thus the natural question emerges: can we use other types of measurements and check whether or not the implications of Eqs. 4 and 5 are apparent in those experiments?

Here we design two new measurement schemes utilizing the energy and force in a capacitor shunting a resistor, where the time-energy uncertainty principle is irrelevant so that we are free from the artifact pointed out by Kleen [17].

3.1. Energy in a shunting capacitor

Consider first the mean energy in a capacitor shunting a resistor. Fig. 3 shows this system, which is a first-order low-pass filter with a single pole at a frequency $f_L = (2\pi RC)^{-1}$. $R$ and $C$ denote resistance and capacitance, respectively.

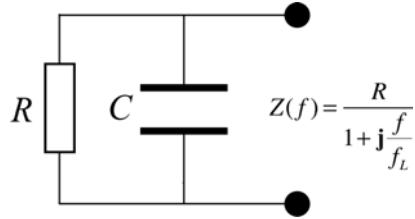

**Figure 3.** Resistor $R$ shunted by a capacitor $C$. $Z(f)$ is the impedance.

The real part of the impedance is given as $\mathrm{Re}[Z(f)] = R(1+f^2 f_L^{-2})^{-1}$ and thus, in accordance with Callen–Welton [3] and Eq. 4, the one-sided power density spectrum $S_{u,C}(f)$ of the voltage on the impedance (and on the capacitor) is

$$S_{u,C}(f) = \frac{4RhfN(f,T)}{1+f^2 f_L^{-2}} + \frac{2Rhf}{1+f^2 f_L^{-2}} \quad , \tag{6}$$

where the first term is classical-physical while the second one is its quantum (zero-point) correction; see Fig. 4.

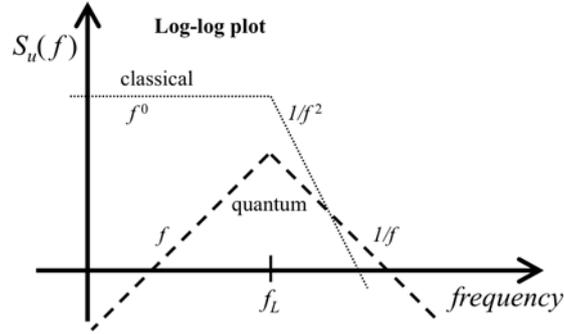

**Figure 4.** Bode plot, with low and high frequency asymptotes, of the classical and quantum (zero-point) component of the power density spectrum of the voltage on the capacitor at finite temperature. The classical Lorentzian spectrum has white and $1/f^2$ spectral regimes. At zero temperature, only the quantum term exists; it is an $f$-noise at low frequencies and converges to $1/f$ at $f > f_L$.

The mean energy in the capacitor is given by

$$\langle E_C \rangle = 0.5 C \langle U_C^2(t) \rangle = 0.5 C \int_0^{f_c} S_{u,C}(f) df ,  \tag{7}$$

where $f_c \gg f_L$ is the cut-off frequency of the transport in the resistor. At near-zero temperature, the classical component $\langle U_{C,c}^2(t) \rangle$ of $\langle U_C^2(t) \rangle$ vanishes, *i.e.*,

$$\lim_{T \to 0} \langle U_{C,c}^2(t) \rangle = \lim_{T \to 0} \left\{ 4Rh \int_0^{f_c} \frac{f[\exp(hf/kT) - 1]^{-1}}{1 + f^2 f_L^{-2}} df \right\} = 0 , \tag{8}$$

but the quantum (zero-point) term remains and is

$$\langle U_{C,q}^2(t) \rangle = \int_0^{f_c} \frac{2hfR}{1 + f^2 f_L^{-2}} df = hR f_L^2 \ln\left(1 + \frac{f_c^2}{f_L^2}\right). \tag{9}$$

Thus the energy in the capacitor, in the zero-temperature approximation, is

$$\langle E_C \rangle = \frac{h}{8\pi^2 RC} \ln\left(1 + 4\pi^2 R^2 C^2 f_c^2\right) . \tag{10}$$

Eq. 10 implies that, by choosing different resistance values, the capacitor is charged up to different mean-energy levels. This energy can be measured by, for example, switching the capacitor between two resistors with different resistance values and evaluating the dissipated heat; see Sec. 4.1 below.

3.2. Force in a capacitor

In a plane circular capacitor, where the distance $x$ between the planes is much smaller than the smallest diameter $d$ of the planes, the attractive force between the planes [21] is given by

$$F = \frac{E_C}{x} \quad . \tag{11}$$

Eqs. 10 and 11 imply that the mean force in the capacitor shunting a resistor (see Fig. 3) is

$$\langle F(x) \rangle = \frac{\langle E_C \rangle}{x} = \frac{1}{x} \frac{h}{8\pi^2 RC(x)} \ln\left[1 + 4\pi^2 R^2 C^2(x) f_c^2\right], \tag{12}$$

where the *x*-dependence of the capacitance is expressed by $C(x) = \varepsilon \varepsilon_0 A / x$. Here *A* is the surface of the planes and $\varepsilon$ is dielectric permeability. Consequently Eq. 12 indicates that, at a given distance *x*, different resistance values result in different forces.

**4. A new approach to assess zero-point Johnson noise: Two "perpetual motion machines"**

The above effects on energy and force in a capacitor could be used to build two different "perpetual motion machines", *provided the zero-point term is available for these kinds of measurements*, as further discussed below. This fact proves that the Fluctuation-Dissipation Theorem (see Equation 4) cannot be correct under general conditions.

4.1. Zero-point noise based "perpetual heat generator"

Fig. 5 delineates a "heat-generator" and comprises an ensemble of *N* units, each containing two different resistors and one capacitor. The capacitors in the units are periodically alternated between the two resistors by centrally controlled switches in a synchronized fashion that makes the relative control energy negligible [21]. The duration $\tau_h$ of the period is chosen to be long enough that the capacitors are sufficiently "thermalized" by the zero-point noise, *i.e.*, $\tau_h \gg \max\{R_1 C, R_2 C\}$. Suppose that $R_1 < R_2$ and that the parameters satisfy $\max\{(4\pi R_i C)^{-1}\} \ll f_c$. Whenever the switch makes the $1 \Rightarrow 2$ transition, the energy difference will then dissipate in the system of $R_2$ resistors as

$$0 < E_h = N \frac{h}{8\pi^2 C} \left[ \frac{\ln\left(1 + 4\pi^2 R_1^2 C^2 f_c^2\right)}{R_1} - \frac{\ln\left(1 + 4\pi^2 R_2^2 C^2 f_c^2\right)}{R_2} \right] \quad . \tag{13}$$

After the reverse $2 \Rightarrow 1$ transition, the capacitors will be recharged by the system of $R_1$ resistors to their higher mean energy level.

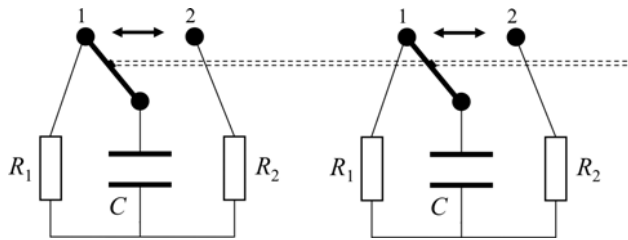

**Figure 5.** Heat-generator based "perpetual motion machine". The coupled switch is periodically alternated between the two states.

Hence the "heat-generator" system pumps energy from the system of $R_1$ resistors to the system of $R_2$ resistors, where this energy is dissipated as heat. The heat can be utilized to drive the switches of this "perpetual motion machine". Such a result violates not only the Second Law of Thermodynamics by its negentropy production in thermal equilibrium, but it also violates the Energy Conservation Law.

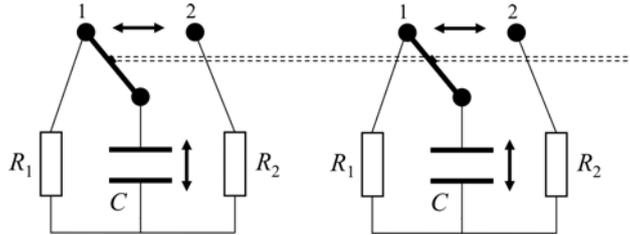

**Figure 6.** A moving-plate capacitor piston based "perpetual motion machine". The coupled switch is periodically alternated between the two states. See also Fig. 7.

4.2. Zero-point noise based "perpetual motion engine"

The second perpetual motion machine is a two-stroke engine; see Fig. 6. This is the zero-point energy version of the two-stroke Johnson noise engine described earlier [21]. The engine has *N* parallel cylinders with elements and parameters identical to those in Fig. 5. The only difference is that the capacitors have a moving plate that acts as a piston. The plates are coupled to a device which moves them in a periodic and synchronized fashion. When the plate separation reaches its nearest and farthest distance limits denoted $x_{min}$ and $x_{max}$ —where the corresponding capacitance values are $C_{max}$ and $C_{min}$, respectively—the switch alternates the driving resistor; see Fig. 7. During contraction, the attractive force between the capacitor plates should be higher than during expansion. Since the force is higher when the capacitor is connected to $R_1$, the driver is $R_1$ and $R_2$ (with $R_1 < R_2$) during contraction and expansion, respectively. At a given distance *x*, the difference in the attractive force between the cases of the capacitor being attached to $R_1$ and $R_2$ is [16]

$$\langle \Delta F(x) \rangle = \frac{1}{x} \frac{h}{8\pi^2 C(x)} *$$
$$* \left\{ \frac{1}{R_1} \ln\left[1 + 4\pi^2 R_1^2 C^2(x) f_c^2\right] - \frac{1}{R_2} \ln\left[1 + 4\pi^2 R_2^2 C^2(x) f_c^2\right] \right\} . \quad (14)$$

At a given value of *x,* the total force difference in *N* cylinders is

$$\Delta F_N(x) = N \langle \Delta F(x) \rangle . \quad (15)$$

The distance changes during contraction and expansion, and therefore the force difference must be integrated over *x*. With $R_1 < R_2$ and at any given plate distance (and corresponding capacitance), the force $N\langle F(x) \rangle$ is stronger during contraction than during expansion; see Fig. 7. During a full cycle, net positive work is executed by the engine according to

$$W = \oint_{x_{\min}, x_{\max}} N \langle F(x) \rangle dx = \int_{x_{\max}}^{x_{\min}} \Delta F_N(x) dx > 0 \ . \tag{16}$$

While this two-stroke engine produces positive work during its whole cycle, a heat-generation effect also sets in for switching at $C_{\max}$, *i.e.*, heat is generated in $R_2$ similarly to what happens in the first perpetual motion machine (see Sec. 4.1).

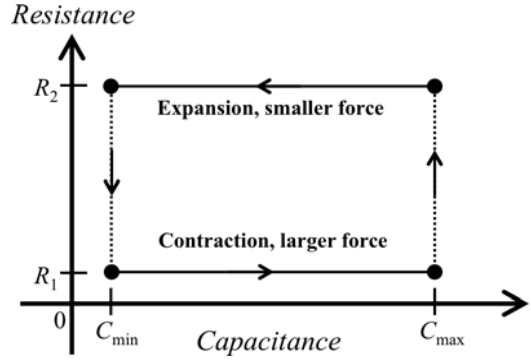

Figure 7. Capacitance–resistance diagram of a two-stroke "perpetual motion engine".

It should be noted that the Casimir effect also implies an attractive force between the capacitor plates. However, the Casimir pressure decays [22] as $x^{-4}$ which implies that the Casimir force, at a fixed capacitance, falls off as $x^{-3}$. At the same time, the force due to the zero-point noise decays as $x^{-1}$. Thus the Casimir effect can always be made negligible in the "perpetual motion machines" via a proper choice of the actual range of $x$ values between the plates during operation.

The two "perpetual motion machines" discussed above explicitly violate not only the Second Law of Thermodynamics but also the Energy Conservation Law. Thus the key assumption underlying their creation, *i.e.*, the Fluctuation-Dissipation Theorem, Eqs. 4 and 5, for the Johnson noise of resistors and impedances, cannot be valid under general conditions.

## 5. Is the Johnson–Nyquist formula valid?

Can we conclude that the zero-point term must be omitted and that the remaining original Johnson–Nyquist formula (Eq. 1) is valid? This is certainly so for the resistor–antenna blackbody radiation scheme described in Sec. 2.2, because any deviation in that situation would violate the Second Law of Thermodynamics: with a deviation in any frequency range, and with proper filters, energy flow could occur under thermal equilibrium between the thermal researvoire containing the resistor and the blackbody radiation.

However, investigating the validity of the Johnson–Nyquist formula by use of our resistor–capacitor circuit leads to a surprise. Suppose that Eq. 1 describes the Johnson noise. It is well known that Eq. 3, in the classical limit of

$$\frac{1}{2\pi RC} = f_L \ll \frac{kT}{h} \ , \tag{17}$$

yields $\langle U_{C,c}^2(t) \rangle = kT/C$ and an ensuing mean energy of $kT/2$ in the capacitor. This is in accordance with Boltzmann's energy equipartition theorem and implies that the Second Law of Thermodynamics is satisfied. But the situation is different in the quantum limit, with

$$\frac{kT}{h} \ll f_L \, , \tag{18}$$

because in the narrow noise-bandwidth caused by the exponential high-frequency cut-off of $N(f,T)$ the voltage noise spectrum of the capacitor is proportional to $R$ so that $\langle U_{C,c}^2(t) \rangle \propto R$, which is evident also from Eqs. 8 and 18. Thus, in the quantum regime, according to Nyquist's old result (Eq. 1), the mean energy in the capacitor varies as

$$\langle E_C \rangle \propto RC \, , \tag{19}$$

which is an inverse scaling compared to the one in the zero-point noise limit; see Eq. 10. The result implies that, in the quantum limit (Eq. 18), the old Johnson–Nyquist formula (Eq. 1) also leads to the "perpetual motion machines" outlined in Sec. 4, except that the direction of the energy flow is opposite. It is also clear that the two energies encapsulated in Eqs. 10 and 19 cannot compensate each other except at a single temperature, which is unimportant when the Second Law of Thermodynamics is violated at other temperatures.

We conclude that not only does the zero-point Johnson noise depend on the external (measurement) circuitry connected to the resistor but, in the quantum limit, Nyquist's old result (Eq. 1) also suffers from the same problem.

## 6. Conclusions and observations

This article is not only a critical assessment of the longstanding debate regarding the zero-point term in the Johnson noise voltage of a resistor but also points out that both Nyquist's result and the Fluctuation-Dissipation Theorem (Eqs. 1 and 4) break down in the quantum limit. Both Nyquist and Callen–Welton were mistaken in their expectation of a general, system-independent formula for a single noise source in the resistor. We strongly believe that the problem does not originate from the lumped (discrete) versus distributed circuit elements situation in the external circuitry, and we observe that we are accompanied in that view by, for example, Nyquist in his classical derivation [2] with a waveguide, Ginsburg–Pitaevskii in their quantum derivation [6] with classical discrete linear circuit elements, and Koch–van Harlingen–Clarke whose experimental analysis [15] employed classical discrete linear and non-linear circuit elements.

A clarification: Zero-point energy does exist, of course! The question here is absolutely different: What is the actual shape of the Johnson noise spectrum in the quantum limit of different types of measurements?

Taking into account the experimental facts, as well as the old and new considerations, leads us to the conclusion that, in the quantum limit, it is impossible to propose a Johnson noise formula that identifies a single, measurement-system-independent (external-circuit-independent) noise source in the resistor to account for the measured noise and its effects. We surmise that this fact is in accordance with the principles of quantum physics, namely that the measurement device interferes with the recorded effect.

Particularly, in the quantum limit, we find that

i. the Johnson–Nyquist formula (Eq. 1) is valid when an antenna is connected to a resistor and Johnson noise is measured by radiation emitted by the antenna,

ii. the Callen–Welton formula (Eq. 4) holds when Johnson noise is measured by a narrow-band quantum device (Josephson junction), and

iii. neither the Johnson–Nyquist formula (Eq. 1) nor the Callen–Welton formula (Eq. 4) is valid when measuring Johnson noise by energy/force in a shunting capacitor.

In the spirit of this Special Issue, three *Unsolved Problems of Noise* stand out:

a) What is the correct Johnson noise equation for case (iii)?

b) Can a generally valid, system-independent set of equations be created if one includes both voltage and current noise generators—instead of one of these as Nyquist [2], Callen–Welton [3] and Kubo *et al*. [5] did)—such as in a theoretical description of classical wideband amplifiers [23]?

c) If satisfying (b) is impossible, it is necessary to involve a separate quantum theory for each circuitry/measurement; are there then particular circuitry/measurement classes wherein the same specific result would hold—in the way that our Eqs. 1 and 4 suggest?

Finally, we observe that the above considerations are not only related to basic science but may also be relevant for technical applications. The issue of the force in a capacitor has potential importance in advanced nanotechnology, where van der Waals and Casimir forces are present [24]. In systems where there is electrical connection between nanostructured conductors that form capacitors, such as coated cantilevers, the zero-point noise would imply forces that could potentially dominate over the van der Waals and Casimir forces. It is however important to note that, whenever the phase coherent length of charge transport becomes larger than the characteristic size of elements, mesoscopic effects of coherent transport must be taken into consideration [25].

Accepted for publication in Journal of Statistical Mechanics: Theory and Experiment
December 20, 2015.  http://arxiv.org/abs/1504.08229
**Acknowledgements**

Valuable discussions with Kyle Sundqvist and Peter Rentzepis are appreciated. We are grateful to an anonymous reviewer who pointed out that considerations about Nyquist's old formula (Sec. 5) should be included in this paper. Though we were aware of these issues, we originally wanted to focus on the zero-point term only and planned to discuss the problems with Nyquist's theory in a follow-up paper. However including these considerations makes the claims in the present paper stronger.